\documentclass[aps,prb,showpacs,superscriptaddress,twocolumn]{revtex4-1}
\usepackage{subfigure}
\usepackage{bm}
\usepackage{amsmath,amsfonts,amssymb,graphicx}
\usepackage{lipsum}

\begin{document}
\title{Proximity Effect in Superconducting-Ferromagnetic Granular Structures}
\author{Hadar Greener}
\email{hadargre@mail.tau.ac.il}
\author{Victor Shelukhin}
\author{Michael Karpovski}
\author{Moshe Goldstein}
\author{Alexander Palevski}
\affiliation{School of Physics and Astronomy, Raymond and Beverly Sackler Faculty of Exact Sciences, Tel Aviv University, Tel Aviv 6997801, Israel}

\begin{abstract}
We examined the proximity effect in granular films made of Pb, a superconductor, and Ni, a ferromagnet, with various compositions. Slow decay of the critical temperature as a function of the relative volume concentration of Ni per sample was demonstrated by our measurements, followed by a saturation of T$_c$.
Using an approximate theoretical description of our granular system in terms of a layered one, we show that our data can only be reasonably fitted by a trilayer model. This indicates the importance of the interplay between different ferromagnetic grains, which should lead to triplet Cooper pairing.

\end{abstract}
\maketitle

%

\section{INTRODUCTION}
Triplet pairing is known as a rather exotic phenomenon, which is expected to be very sensitive to disorder and spin orbital interaction~\cite{ref39larkin}, making its observation very challenging.
Nevertheless, evidence for triplet pairing was found in a few systems, including superfluid $^{3}$He,\cite{leggethe3} and the superconducting perovskite Sr$_{2}$RuO$_{4}$.\cite{ref38mackenzie}
The possibility of generating a triplet superconducting component due to the proximity effect between singlet superconductors (S) and non-homogeneous ferromagnets (F) is among the many fascinating proximity-affected phenomena studied in such hybrid systems. 

Theoretically, the suppression of superconductivity in SF layered structures is caused mainly by the exchange interaction of the ferromagnet: the exchange field in a typical ferromagnet is by orders of magnitude larger than the BCS energy gap, interpreted here as the Cooper binding energy.
As a result, singlet Cooper pairs, composed of electrons of opposite spin orientations, cannot penetrate into the F layers beyond the typically short magnetic length. To put it differently, singlet Cooper pairs are easily destroyed, since the spins of the electrons cannot be antiparallel in the presence of the ferromagnetic exchange field. 
The situation might be different if the spins of the superconducting
Cooper pairs were parallel to each other. Clearly, such triplet Cooper pairs would not be sensitive to the ferromagnetic exchange field, and some sort of  coexistence of superconductivity and ferromagnetism would become possible. Thus, the critical temperature of the system would decay more slowly and might even display reentrant behavior.

In line with this, in SF multilayers, it has been shown theoretically~\cite{ref36fominov} that the Cooper pair-breaking effect can depend on the relative orientation of the magnetic moments in the F layers. 
In particular, the critical temperature $T_{c}$ exhibits reentrant behavior and a local maximum as function of the F layers' thickness. Although in these systems only a spin-singlet pairing component exists in the bulk superconductor, the interplay between ferromagnetism and superconductivity gives rise to the emergence of long-range spin triplet components in FSF structures, when the magnetic moments in each of the F layers are non-collinear~\cite{ref37champel}. It has been found experimentally by Shelukhin \emph{et al.}~\cite{shelukhin2006}, that these layered structures exhibit non-monotonic behavior of $T_{c}$ as a function of the thickness $d_f$ of the ferromagnetic layer, in compliance with the predicted formation of a $\pi$-phase junction, which occurs when a ferromagnet is sandwiched between two superconductors with order parameters of opposite sign~\cite{bulaevskiiref29, buzdinref8,buzdinref9,bergeretref10}. Similar results have also been shown \cite{PhysRevB.44.759,PhysRevLett.74.314} for the short range spin triplet component.

In this paper, we report studies of the critical temperature variations as a function of the relative volume concentration of Ni, a ferromagnet with a Curie temperature of $627$~K, in hybrid
Ni-Pb granular structures, where Pb is a spin-singlet superconductor with a bulk critical temperature of $T_{c,0}=7.2$~K.
We find a relatively slow decay of T$_c$ followed by saturation as the Ni concentration is increased.
Motivated by previous research conducted on granular mixtures of a superconductor and a non-magnetic normal metal~\cite{sternfeld2005}, we compared our measurements with the theories developed for SF bilayers~\cite{ref36fominov} and FSF trilayer models~\cite{lofwander,FominovGoolubovKupriyanov}, with the layer thickness ratio replaced by the relative volume concentrations $P[$Ni$]/(1-P[$Ni$])$. Between the bilayer and trilayer models, we find that the measured dependence of the critical temperature $T_c$ on this ratio can be reasonably explained only by invoking the trilayer model, where triplet Cooper pairs can come into play.

\section{SAMPLE PREPARATION}

The systems studied were two dimensional random
Ni-Pb granular mixtures, with various ferromagnetic compositions, deposited on
a GaAs substrate. Two sets of 8 samples were prepared separately,
using a typical photolithography liftoff technique in
a four terminal Hall bar configuration, with gold ohmic contacts evaporated
using an \emph{e}-gun. The variation of the Ni relative volume concentrations
was achieved by a co-sputtering technique, using a specially designed
shutter, which exposed the samples in sequence to a constant Pb sputtering
power, and simultaneously, to an increasing Ni sputtering power. The overall
height of the samples did not exceed $500~\mathrm{\AA}$, in compliance with the
\emph{Cooper limit} \cite{CooperLimit1,CooperLimit2,CooperLimit3}, where the grain size is not much larger than the superconducting coherence length (see Subsec.~\ref{subsec:xis} below). A $200~\mathrm{\AA}$ protective layer of Ge was evaporated \emph{in
situ} on top of the films to prevent oxidization. Energy-dispersive
spectroscopy (EDS) X-ray analysis was performed in order to determine
the chemical composition and the relative volume concentration $P[$Ni$]/(1-P[$Ni$])$
for each sample. Due to inhomogeneities of grain sizes and distribution in the samples, seen in Figure \ref{fig:ESEM-backscattered-electron}, the volume
concentration was calculated as the mean of five different points
measured across the sample, while the error was calculated according
to the standard deviation of these measurements. Transport measurements
were performed in a $^{4}$He cryostat, with a temperature range of down to 1.5~K, using a standard AC lock-in technique.

\section{EXPERIMENTAL DATA}

\begin{figure}
\includegraphics[scale=0.3]{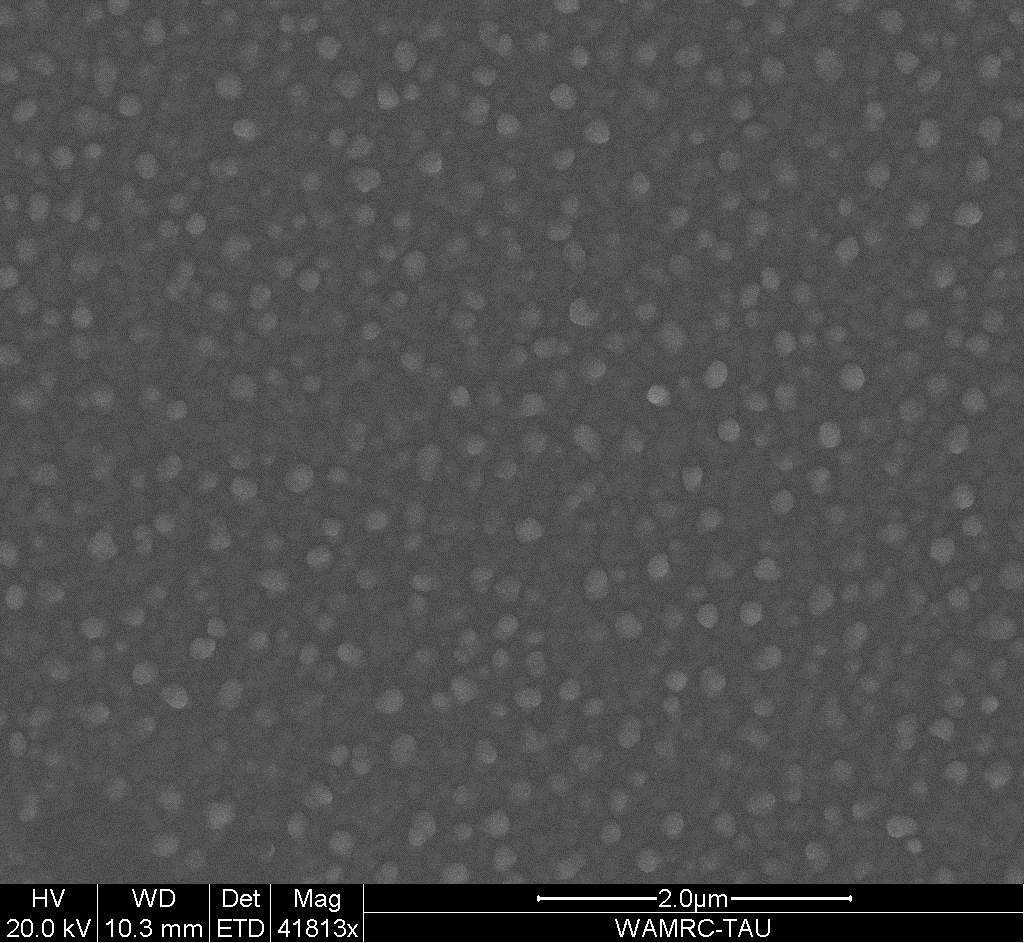}\centering
\protect\caption{ESEM backscattered electron micrograph of a typical Pb-Ni sample with  $P[$Ni$]/(1-P[$Ni$])=0.241$.
Bright regions correspond to higher atomic numbers, and therefore correspond to
Pb grains.
\label{fig:ESEM-backscattered-electron}}
\end{figure}

\begin{figure}[h!]
\begin{minipage}{\columnwidth}
\includegraphics[scale=0.22]{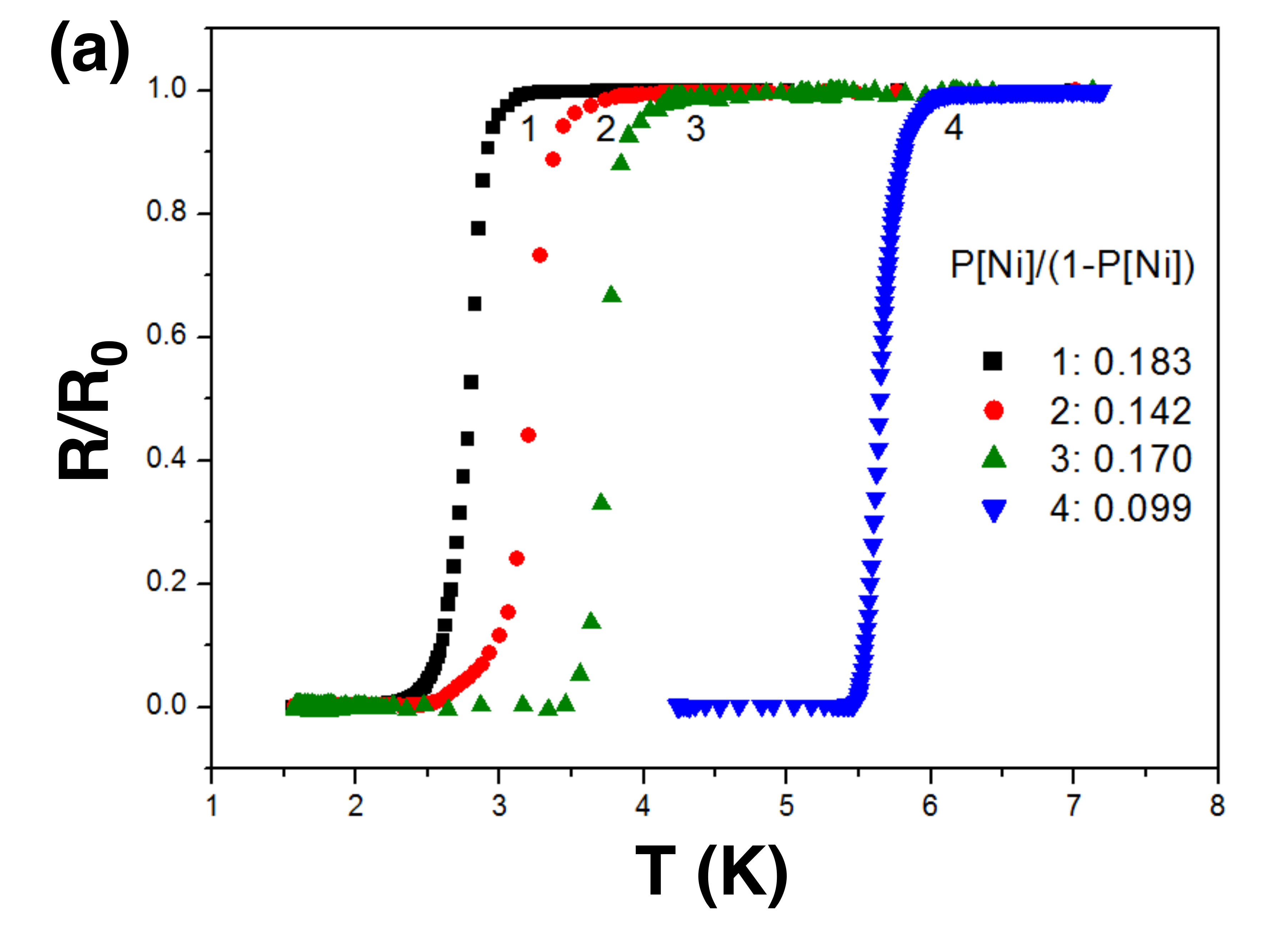}
\end{minipage}
\begin{minipage}{\columnwidth}
\includegraphics[scale=0.22]{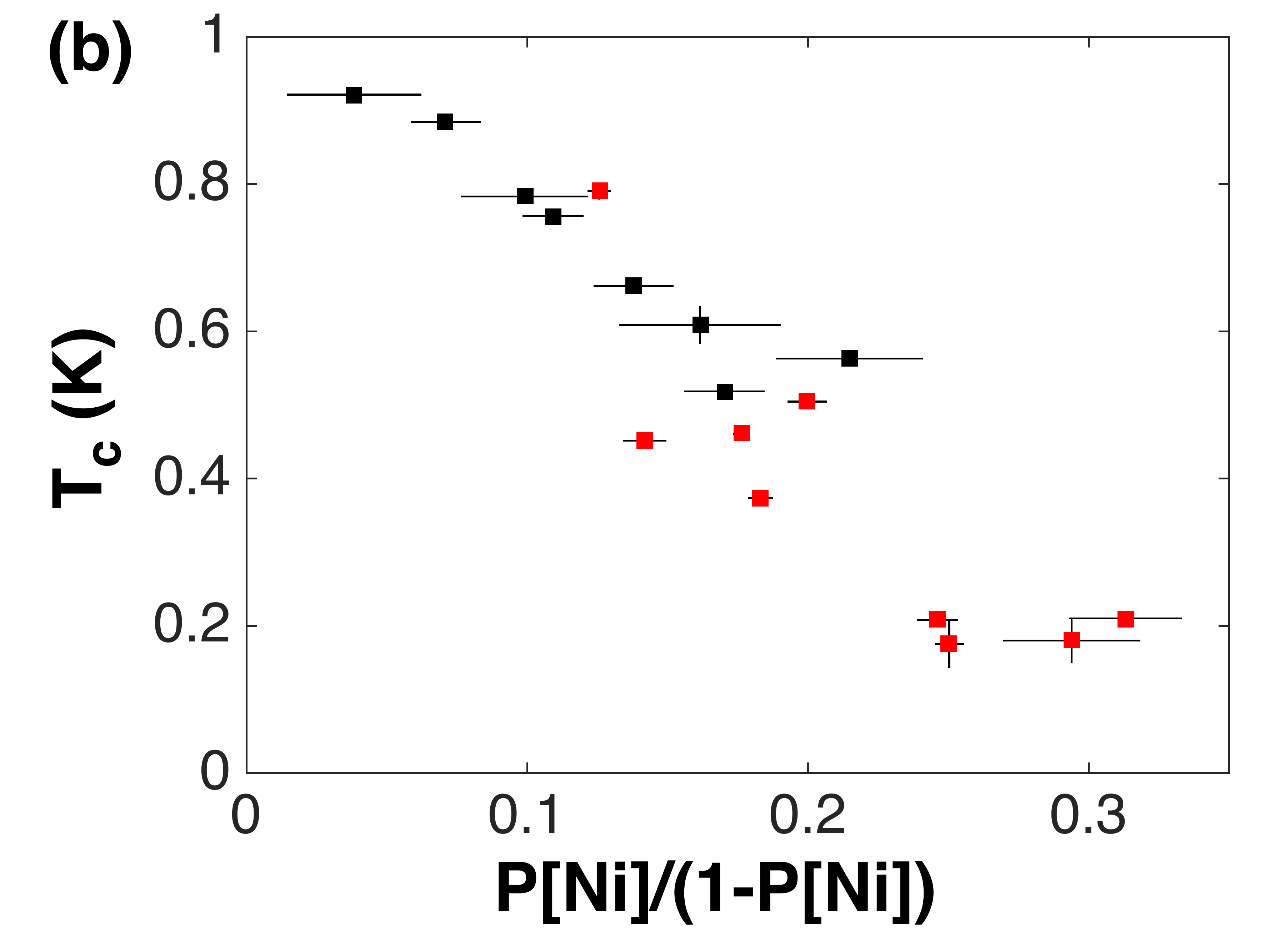}
\end{minipage}
\protect\caption{(Color online) (a) Resistance (normalized by the normal state resistance) as a function of temperature, for typical granular Pb-Ni
samples. (b) Critical temperature measured as a function of the relative Ni volume concentration per sample for two data sets. Black: First batch of samples. Red: Second batch of samples. 
\label{fig:Tcmeasurements}}
\end{figure}

\begin{figure}
\includegraphics[scale=0.22]{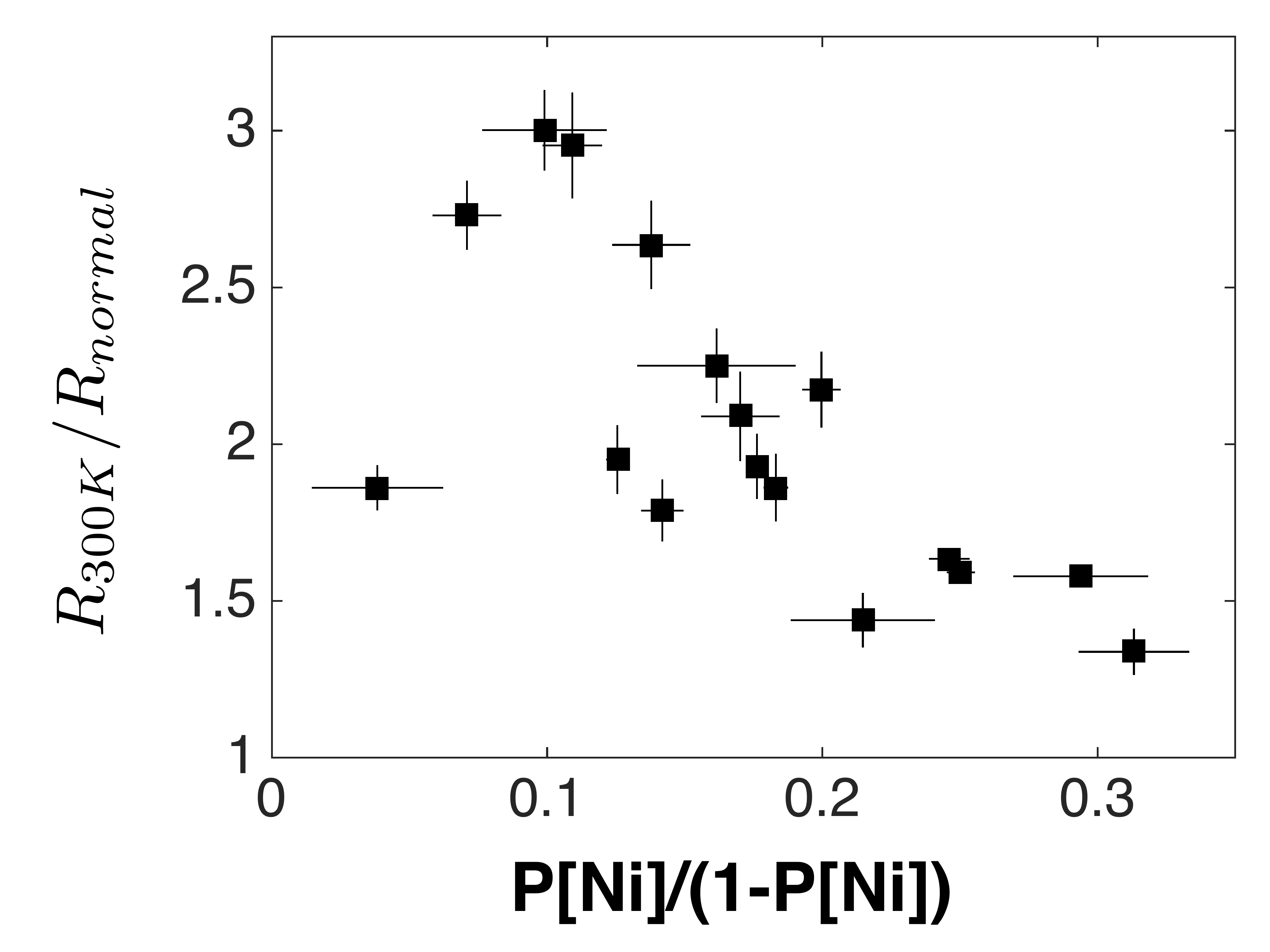}
\protect\caption{(Color online) Resistance at 300K normalized by the normal state resistance (measured just above the superconducting transition) as a function of relative Ni volume concentration.
\label{fig:Rat300K}}
\end{figure}

\begin{figure}
\includegraphics[scale=0.22]{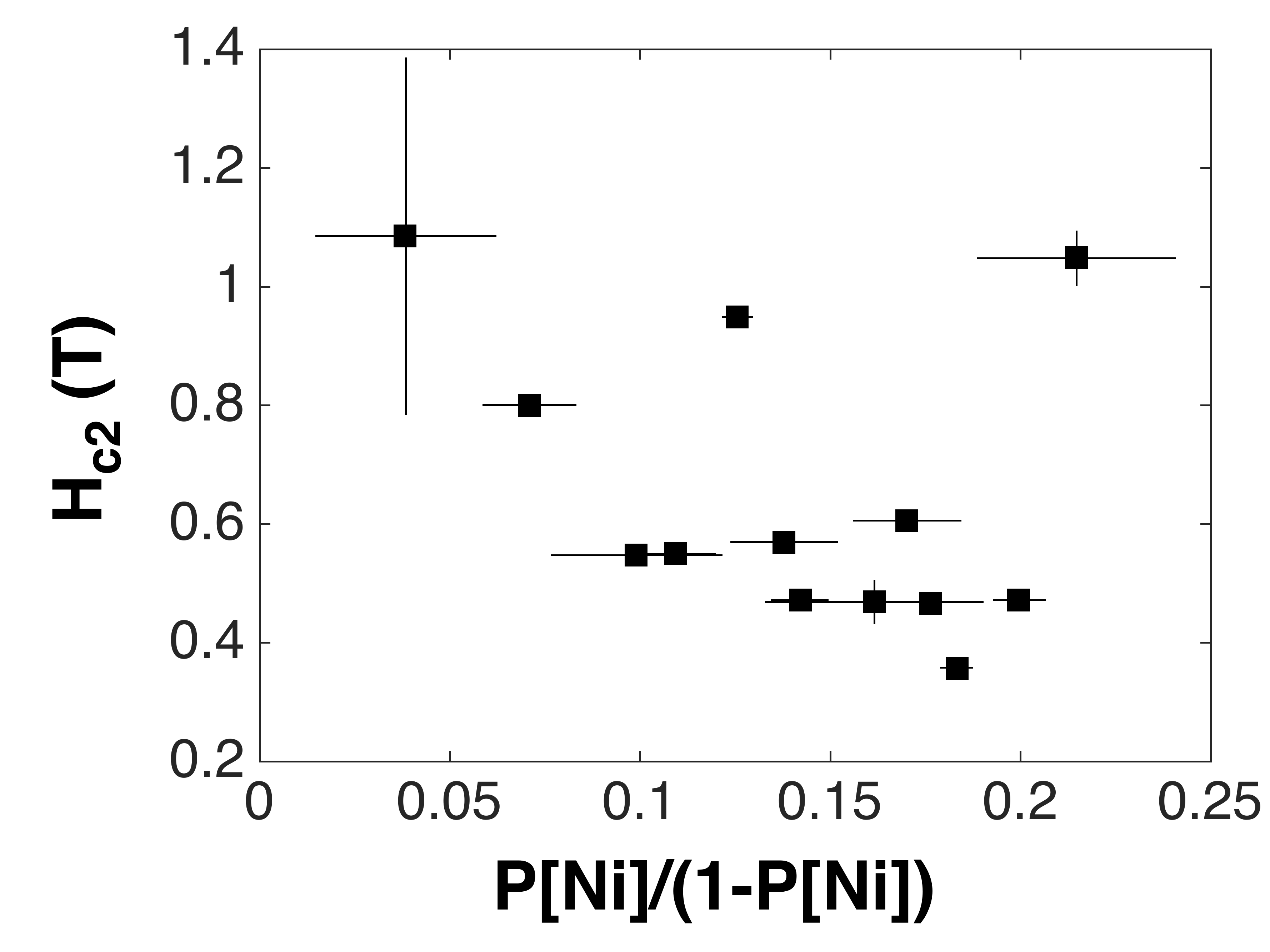}
\protect\caption{(Color online) Measured magnetic field at $T=1.5$~K as a function of the relative Ni volume concentration per sample.
\label{fig:Hcmeasurements}}
\end{figure}

Resistivity versus temperature measurements
were carried out at cryogenic temperatures down to 1.5~K for two different sets of samples with relative
volume concentrations varying between $0.038<P[$Ni$]/(1-P[$Ni$])<0.313$.
A superconductivity phase transition was observed for all of the samples,
while a full phase transition, defined by the experimental observation of zero
resistivity, was measured for all samples with $P[$Ni$]/(1-P[$Ni$])<0.246$. Typical transition curves for four of the samples are shown in Figure \ref{fig:Tcmeasurements}(a). For each sample, the transition temperature $T_{c}$ was defined
as the temperature for which the resistance reached half of its value at the normal state above the critical temperature of bulk Pb, $T_{c,0}=7.2$~K. Figure \ref{fig:Tcmeasurements}(b) presents the critical temperature measured for each sample of the two different sets, shown as black and red data points. The large horizontal error bars in the data points are caused by inhomogeneities of the given sample, originating in the co-sputtering
method. These inhomogeneities were observed in the elemental analysis
of the different samples. Furthermore, it was apparent from backscattered
environmental scanning electron microscope (ESEM) micrographs of the
samples (see Figure \ref{fig:ESEM-backscattered-electron}), in which
only large Pb grains could be resolved, that the grains vary in size
(20-200~nm). This lead us to believe that our structures might exhibit
bi-modal behavior, with large Pb grains that are far apart (not percolating), and much smaller Pb and Ni grains. This means that $T_{c}$ changes locally, and that the $T_{c}$ we measured is the highest of a cluster of grains with similar parameters which percolate through the structure. This will be taken into account in the theoretical analysis of our data.
Additionally, given the range of superconducting coherence lengths,
calculated using our measurements of the resistance at the normal state above the transition for each of the samples and the second critical field H$_{c2}$ measured only for samples that exhibited a full phase transition (see Figs.~\ref{fig:Rat300K} and~\ref{fig:Hcmeasurements})
according to the Ginzburg-Landau theory in the dirty limit~\cite{tinkham},
it is evident that most samples are in the \emph{Cooper limit} \cite{CooperLimit1, CooperLimit2, CooperLimit3}, i.e., the grains are not too large as compared to the superconducting coherence length, as will be explained in the next section.
 
The two different sets of data, shown in Figure \ref{fig:Tcmeasurements}(b), feature non-monotonic behavior, with local minima and a maxima that occur around $P[$Ni$]/(1-P[$Ni$]) \approx 0.22$. However, taking into account the error bars in the relative Ni volume concentration and the measured transition temperatures, the claim of this observation cannot be decisive. If these non-monotonic variations were conclusive, that would clearly indicate the formation of $\pi$ junctions resulting from the contribution of spin-triplet Cooper pairs~\cite{shelukhin2006}. Therefore, in our analysis below we employed two different models: a trilayer model, which takes into account the spin-triplet component~\cite{ref36fominov}, and a bilayer model that does not\cite{lofwander}.

\section{ANALYSIS}
\subsection{ESTIMATION OF THE SUPERCONDUCTING COHERENCE LENGTH}
\label{subsec:xis}
\begin{figure} 
\includegraphics[scale=0.22]{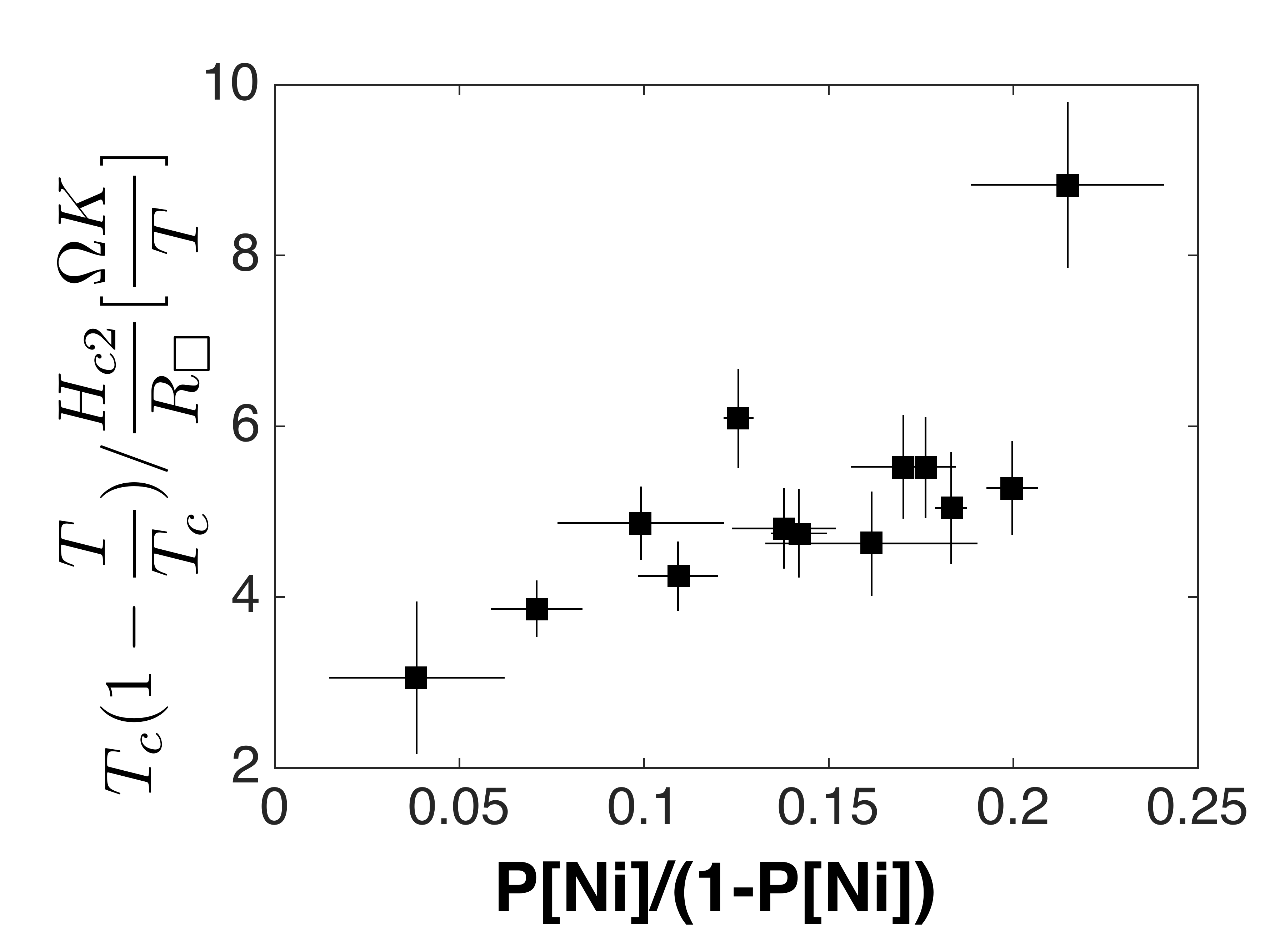}\centering
\protect\caption{(Color online) Comparison of the critical field (at $T=1.5$~K) and normal state resistance data (Figs.~\ref{fig:Rat300K} and ~\ref{fig:Hcmeasurements}) to the Ginzburg-Landau theory, according to which the ratio $T_{c}(1-T/T_{c})/(H_{c2}/R_{\square})$ should be constant. One could easily notice that for larger values of relative volume concentrations, in the range 0.1--0.2, the data is concentrated around an average value of $5.1\pm0.4$~$\Omega$ K/T, with a correlation coefficient of $R^{2} =  0.54$, which implies a weak correlation.
\label{fig:GLtheory}}
\end{figure}

According to the Ginzburg-Landau theory, the critical magnetic field is given by $H_{c2}\sim\ \Phi_{0} / 2\pi \xi^2$, where $\Phi_0$ is the flux quantum, and $\xi$ is the Ginzburg-Landau coherence length. Therefore, in the \emph{dirty limit}, where $\xi \sim \sqrt{\xi_{0} \ell}/(1-T/T_{c})$, with $\xi_0$ the BCS coherence length and $\ell$ the mean free path~\cite{tinkham}, one would expect $T_{c}(1-T/T_{c})/(H_{c2}/R_{\square})$ to attain a constant value, where $R_{\square}$ is the sheet resistance per square of each sample. Figure \ref{fig:GLtheory} shows this ratio plotted for each measured sample. One can immediately notice an almost constant trend line, averaging at $5.1\pm0.4$~$\Omega$ K/T,
for all samples with $P[$Ni$]/(1-P[$Ni$])>0.1$, with the exception of the sample with the highest Ni concentration. 
Thus, from the critical magnetic field shown in Figure \ref{fig:Hcmeasurements}, we obtain an average coherence length of $\xi_{s}=200~\mathrm{\AA}$ for our samples.

The success of the Ginzburg-Landau description of the behavior of H$_{c2}$ indicates that most of our samples are in the \emph{Cooper limit} \cite{CooperLimit1, CooperLimit2, CooperLimit3}, i.e., the size of the typical grains is comparable to the Ginzburg-Landau coherence length.
This also offers a possible explanation for the deviations from this description for lower relative volume concentrations of Ni, where Pb grains tend to be larger, and thus are no longer in the strict \emph{Cooper limit}.

\subsection{CRITICAL TEMPERATURE FOR MULTILAYERED STRUCTURES} 
While the layered SF structures were well studied~\cite{ref36fominov, ref37champel, lofwander}, theoretical modeling of granular SF mixtures is lacking. 
Following successful attempts made in the past to describe granular normal metal-superconductor (NS) mixtures as multilayered structures~\cite{shelukhin2006}, we attempt a similar approximate description for the current system. Its semiquatitative validity is supported by the fact that most of our samples are in the \emph{Cooper limit}, as discussed above.
We will therefore compared our data to two existing models of layered SF structures: a bilayer and a trilayer. The latter model involves more than one F layer. In the generic case, when the magnetization directions of these layers are not collinear (which definitely fits our systems), it inherently invokes the contribution of a superconducting triplet component. The possibility of generating a triplet superconducting component due to the proximity effect between singlet superconductors and non-homogeneous ferromagnets is interesting, since such a component would not be sensitive to the ferromagnetic exchange field, thus the critical temperature would decay more slowly, or even display reentrant behavior with a peak as function of the F layer thickness.

L\"ofwander \emph{et al.}~\cite{lofwander} developed an effective
method to numerically calculate $T_{c}$ in diffusive hybrid SF layered structures,
generalizing Fominov's treatment~\cite{ref36fominov} to the case of asymmetric multilayers.
Let us briefly describe their work, concentrating on the simpler case of a trilayer, consisting of a superconductor sandwiched between two ferromagnets (FSF).

In diffusive structures the Green function is nearly isotropic.
Near $T_{c}$ the energy gap is small, and hence the standard Green function is approximately equal to the its normal-state value, while the anomalous Green's function is proportional to the energy gap. 

In this case, one could use Usadel's linearized diffusion equation for the anomalous Green's function, taking into account both the spin singlet and spin triplet components, $f=(f_s+\vec{\sigma}\cdot\vec{f}_t)i\sigma_{y}$, where $\vec{\sigma}$ are the Pauli matrices.
Assuming
that the spatial dependence in the structure is only along the normal to the interface (the $x$ axis), the Usadel equations take the form:
\begin{eqnarray}
\left(\hbar D(x)\partial_{x}^{2}-2|\epsilon_{n}|\right)f_s(x) & = & -2\pi\Delta(x)+2i\text{sgn}(\epsilon_{n})\vec{J}(x)\cdot\vec{f}_t(x), \nonumber \\
\left(\hbar D(x)\partial_{x}^{2}-2|\epsilon_{n}|\right)\vec{f}_t(x) & = & 2i\text{sgn}(\epsilon_{n})\vec{J}(x)f_{s}(x),
\label{eq:UsadelEquations}
\end{eqnarray}
where $\hbar$ is Planck's constant, $D(x)$ denotes the diffusion constant (taking values $D_s$ and $D_{fi}$ in the S and the two F layers, $i=1,2$, respectively), and $\epsilon_{n}=\pi k_B T(2n+1)$ are fermionic Matsubara energies ($k_B$ is Boltzmann's constant and $n$ is an integer). $\vec{J}(x)$ is nonzero in the ferromagnetic layers, where it is a constant $J_i$ ($i=1,2$) equal to the ferromagnetic exchange energy normalized by the bulk superconductor transition temperature $T_{c0}$.
For simplicity, in the following we will take $D_{f1}=D_{f2}=D_f$ and $J_1=J_2=J$.
Additionally, the gap $\Delta(x)$ is nonzero only in the superconducting regions, where it satisfies the self-consistency equation:
\begin{equation}
\Delta(x) \ln \frac{T}{T_{c0}} = T\sum_{\epsilon_{n}}\left(f_{s}\left(\epsilon_{n},x\right)-\frac{\pi\Delta(x)}{|\epsilon_{n}|}\right).
\end{equation}
These equations are supplemented by the following boundary conditions, that depend
on the coherence lengths in each layer, $\xi_{s}=\sqrt{\hbar D_{s}/2\pi k_B T_{c0}}$
and $\xi_{f}=\sqrt{\hbar D_{f}/2\pi k_B T_{c0}}$: 
\begin{eqnarray}
\gamma\xi_{f} \partial_x f(x_{f}) & = & \xi_{s} \partial f(x_{s}), \nonumber \\
\gamma_{b}\xi_{f} \partial f(x_{f}) & = & \pm\left[f(x_{s})-f(x_{f})\right], 
\label{eq:boundaryconditions}
\end{eqnarray}
where $x_s$ and $x_f$ denote the superconducting and ferromagnetic sides of each SF boundary, and the sign in the second equation is positive at an FS interface and negative in the SF case.
Here, $\gamma=\rho_{s}\xi_{s}/\rho_{f}\xi_{f}$ and $\gamma_{b}=R_{b}A/\rho_{f}\xi_{f}$
are proportional, respectively, to the ratio between the normal-state resistivities of the S and F materials ($\rho_{s}$ and $\rho_{f}$) and the normalized
resistance of the SF boundary ($R_{b}$ per boundary area).

In addition, $\partial_x f = 0$ at the outer boundaries.
One could easily show that for $\vec{f}_{t}\Vert\vec{J}$, $f_t$ becomes a scalar, and the above equations become equivalent to the equations and boundary
conditions developed by Fominov for an SF bilayer~\cite{ref36fominov}.
The analytical and numerical solution of these equations~\cite{lofwander} is summarized in the Appendix.

Let us now discuss the values of the parameters appearing in these equations. In our system the superconducting critical temperature of bulk Pb is $T_{c0}=7.2$~K, and the exchange interaction of Ni is $E_{ex}=200$~meV, as extracted from SF multilayer measurements~\cite{exchangefieldref}, leading to $J=E_{ex}/k_B T_{c0}=322$. Since the samples were grown as a mixture \emph{in situ}, there should not be any barrier associated with the oxide at the inter-grain boundary. Thus the transparency for the trilayer model between the two materials should be high, and we set $\gamma_{b}=0$.

We now turn to $\gamma=\rho_{s}\xi_{s}/\rho_{f}\xi_{f}$,
which is more complicated to estimate.
From their definitions, $\xi_{s}/\xi_{f} = \sqrt{D_{s}/D_{f}}$. Hence, since $D \propto 1/ \rho N$, where $N$ is the density of states, we find $\gamma=\sqrt{\rho_{s}N_{f}/\rho_{f}N_{s}}$.

Now, the normal state resistivities one needs to plug into this relation
are those of the intra-grain material at low temperatures in the normal state.
We can estimate those using values of the room temperature resistivities appearing in the literature. Here one must recall that room temperature resistivities are given for pure materials, where scattering, and therefore the mean free path, are governed by electron-phonon interactions. At low temperatures and in a granular material the mean free path for electronic scattering will be set by the grain size.

Therefore, we estimate the normal resistivity of Pb using the room temperature value, normalized by the room temperature to low-temperature resistivity ratio factor, taken from our measurements in Figure \ref{fig:Rat300K}: $\rho_{s} = \rho_{s,Room}/RR$, where $\rho_{s,Room} = 22 \cdot 10^{-8} \Omega m$ \cite{ashcroftandmermin} and $RR \approx 3$.

Our experience with Ni films grown in similar conditions give an estimate of $\rho_{f} = 5 \cdot 10^{-7} \Omega m$ for the low temperature resistivity\cite{NiRho} of Ni. For the density if states of Ni we will use the value for the s band, which dominates transport. Taking the Fermi energy of s-electrons in Ni to be $E_{f}=4.35eV$, and the effective mass to be $1.5$ times the electron mass \cite{NiFermi}, we find a density of states of $N_{f} = 0.31 (eV atom) ^{-1}$. Plugging in the density of states calculated from the superconducting critical temperature\cite{DOSPb} of Pb, $N_{s} = 0.276 (eV atom)^{-1}$, we find that a reasonable value of $\gamma$ is around $0.4$.

In the following fits we use as a fitting parameter the value of $\gamma$ for low Ni concentration, and then adjust it for other values of $P$[Ni] using the relation $\gamma \approx \sqrt{\rho_s/\rho_f}$, assuming that $\rho_f$ is approximately constant between the samples (since the Ni grain size does not seem to vary), and using the data from Figure \ref{fig:Rat300K} to track the changes in $\rho_s$.

Finally we discuss the different length scales appearing in the theoretical model.
The coherence length $\xi_s$ in Pb is estimated to be 200~\AA  for the higher Ni concentrations, as found in Subsec.~\ref{subsec:xis} above. $\xi_f$ for high Ni concentration is then determined by $\xi_f/\xi_s \approx \sqrt{\rho_s/\rho_f} \approx \gamma$, and it is assumed that it is does not vary much with $P$[Ni], due to the approximate constancy of the Ni grain sizes.
The total width of the F layers is set to $P$[Ni] times 500~\AA\ (the thickness of our granular films); For a bilayer this is just $d_f$, while for the FSF trilayer this determines $d_{f1}+d_{f2}$, while the ratio $d_{f1}/d_{f2}$ is taken as 2 (the results were found to be insensitive to the exact value of this ratio).
As for $d_s$, it should represent the Pb grain size, and thus $d_s \sim 1/\rho_s$ according to the above considerations. The results only depend on $d_s/\xi_s \sim 1/\sqrt{\rho_s}$. We use its value for high Ni concentration (where the Pb grain size is about 1000~\AA\ and $\xi_s$ is close to 200~\AA, so their ratio should be between 4--6) as a fitting parameter, and adjust it to other values of $P$[Ni] using its dependence on $\rho_s$ and the data from Figure \ref{fig:Rat300K}.
In the FSF case we take the angle between the magnetizations of the two F layers as $\theta = \pi/2$, after verifying that our results are not too sensitive to its precise value.

\begin{figure}
\begin{minipage}{\columnwidth}
\includegraphics[scale=0.22]{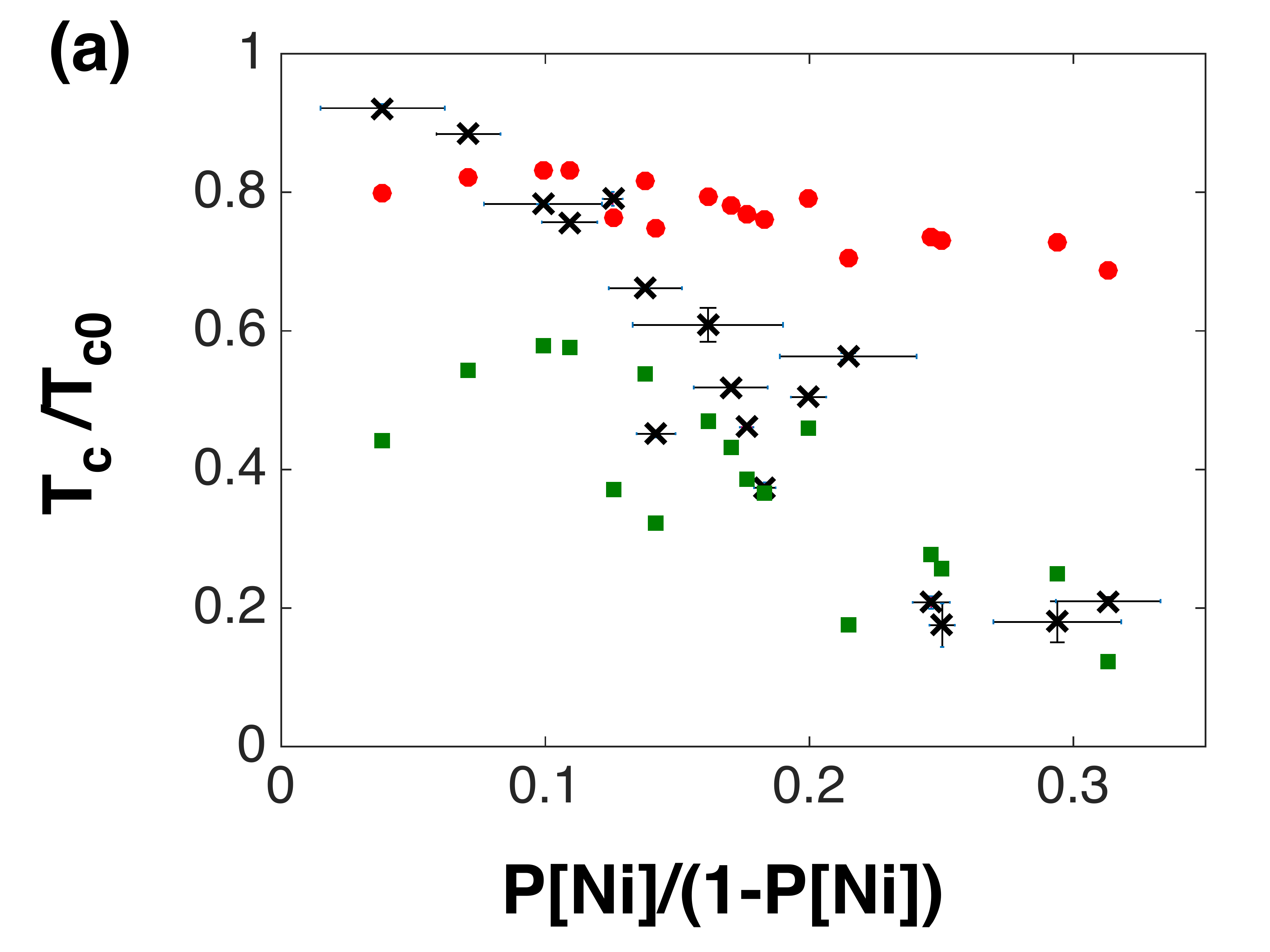}\centering
\end{minipage}
\begin{minipage}{\columnwidth}
\includegraphics[scale=0.22]{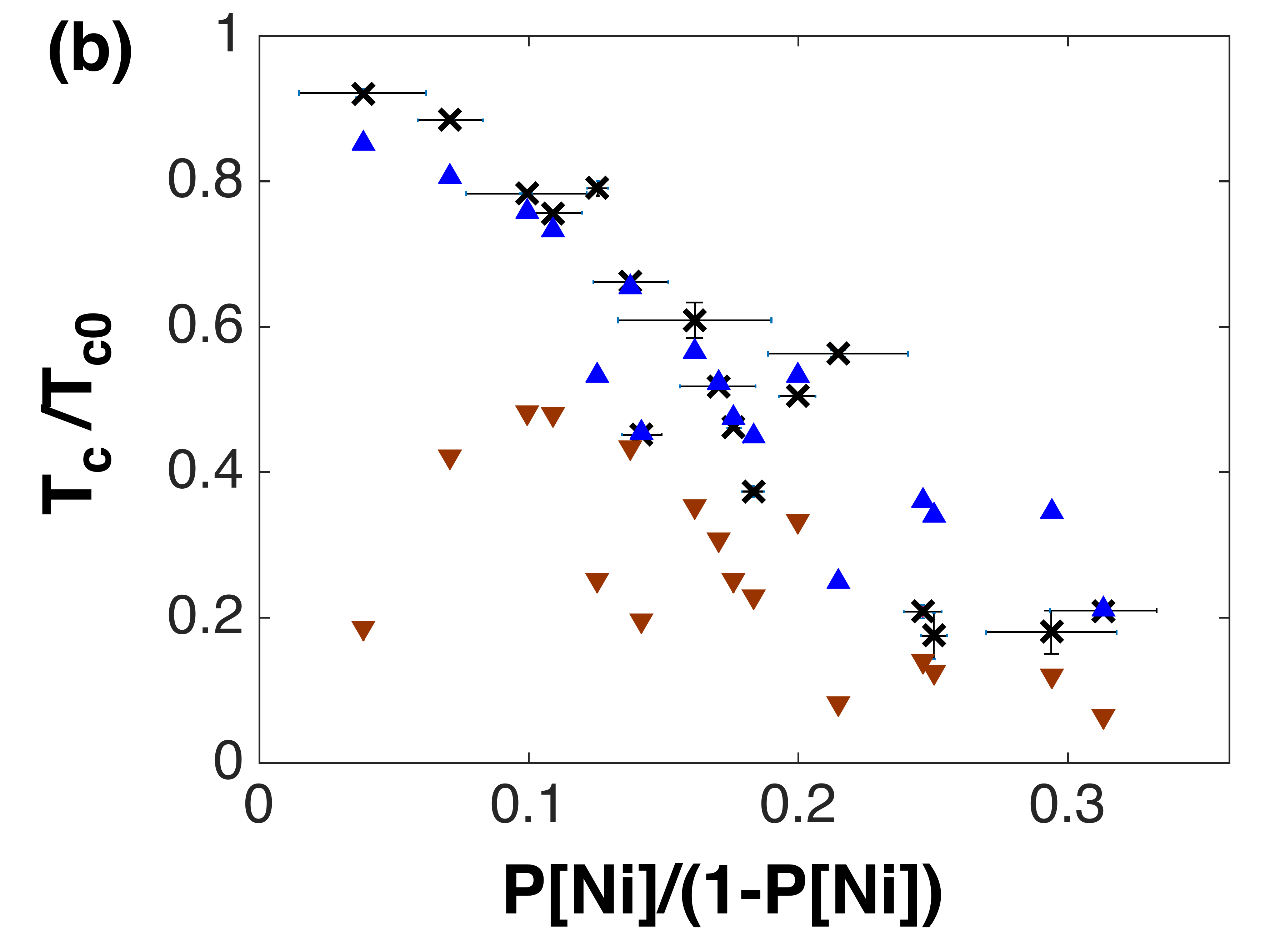}
\end{minipage}
\protect\caption{(Color online) (a) Black crosses: $T_{c}/T_{c0}$ as a function of the relative Ni volume concentration in the experiment (as in Figure \ref{fig:Tcmeasurements}). Red circles: Theoretical calculation of $T_{c}/T_{c0}$ as a function of the relative Ni volume concentration within the SF bilayer model~\cite{ref36fominov}, with $\gamma$ ranging from 0.34 for low Ni concentrations to 0.51 for high Ni concentrations, and $d_s/\xi_s$ ranging from 8 for low Ni concentrations to 5.4 for high Ni concentrations. Green squares: The same as the red circles, but with $d_s/\xi_s$ ranging from 2.6 for low Ni concentrations to 4 for high Ni concentrations.  (see the text for details and other parameters) (b) Black crosses: Same as (a). Blue triangles up: Theoretical calculation of $T_{c}/T_{c0}$ as a function of the relative Ni volume concentration within the SF bilayer model~\cite{ref36fominov}, with $J=150$. Here $\gamma$ and $d_s/\xi_s$ are as red circles in (a). Brown triangles down: Same as the previous case, but with $J=1000$.
\label{fig:BilayerTheory}}
\end{figure}

\begin{figure}[h!]
\includegraphics[scale=0.22]{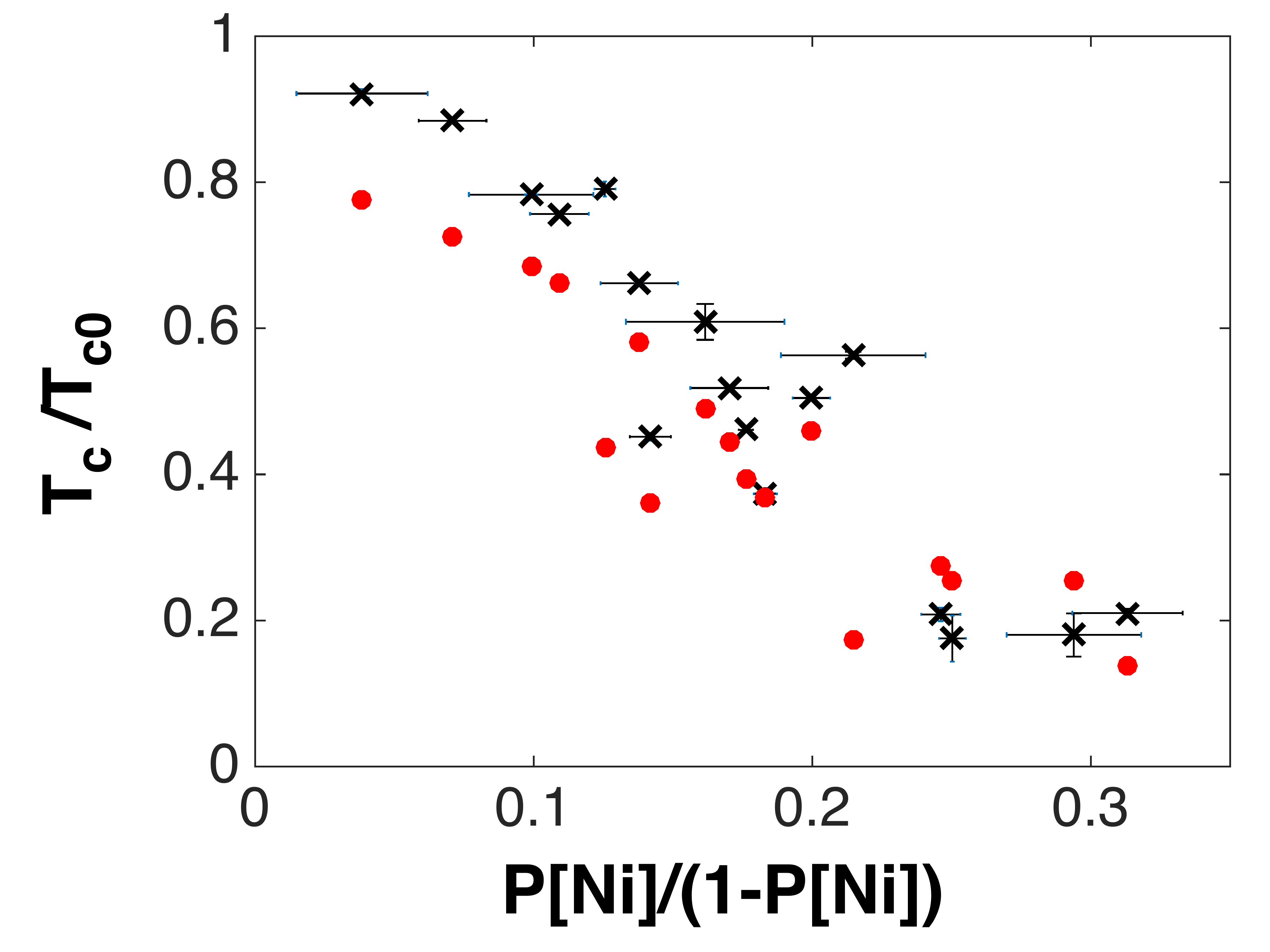}\centering
\protect\caption{(Color online) Black crosses: $T_{c}/T_{c0}$ as a function of the relative Ni volume concentration in the experiment (as in Figure \ref{fig:Tcmeasurements}). Red circles: Theoretical calculation of $T_{c}/T_{c0}$ as a function of the relative Ni volume concentration within the FSF trilayer model~\cite{ref36fominov}, with $\gamma$ ranging from 0.34 for low Ni concentrations and 0.51 for high Ni concentrations, and $d_s/\xi_s$ ranging from 5.4  for high Ni concentrations to 8 for low Ni concentrations (see the text for details and other parameters). 
\label{fig:TrilayerTheory}}
\end{figure}

To conclude, we are left with only two fitting parameters: (a) the values of $\gamma$ for low Ni concentration, which can vary between 0.3--1; (b) the size of $d_s/\xi_s$ for high Ni concentrations, which can vary between 4--6.
With those parameter values we have tried to fit our data to the bilayer and trilayer theories. 
As can be seen in Figs. \ref{fig:BilayerTheory} and \ref{fig:TrilayerTheory}, only the latter can reasonably fit the data but not the former. The best fit was achieved for a trilayer model with values of $\gamma$ ranging from 0.34 for low Ni concentrations to 0.51 for high concentrations, and $d_{s}/\xi_{s}$ ranging from 5.4 for high Ni concentrations to 8 for low concentrations. The same parameters plugged into the bilayer model did not result in a good fit to our data. In this model, the change of $T_{c}$ with $P[$Ni$]/(1-P[$Ni$])$ is weaker than that of the trilayer model.

Tweaking the values of $d_{s}/\xi_{s}$ in this case improved the fit for high Ni concentrations, but failed to simultaneously agree with our data for low Ni concentrations. Figure \ref{fig:BilayerTheory}(b) shows examples of calculations of $T_{c}$ vs. $P[$Ni$]/(1-P[$Ni$])$ within the bilayer model for different limits of $J = E_{ex}/k_{B}T_{c0}$ ($J=150$ and $J=1000$). One sees that a reasonable fit is achieved for an unreasonably low value of $J$, which is not consistent with previous measurements \cite{exchangefieldref}, since $J=150$ is equivalent to an exchange energy of around 90meV.  
 This is an evidence for the importance of the interplay of multiple Ni grains in our samples, which (in the generic case, where their magnetization vectors are not collinear), should result in spin-triplet Cooper pairs.

Non-monotonic behavior is  expected in multilayered SF structures with more than one S layer, in which a $\pi$ Josephson junction may occur; this effect can be important in the variation of $T_{c}$, as was verified in past experiments~\cite{shelukhin2006}. In our granular systems, which are a mixture of grains with different sizes, it seems likely that such an effect will average out, and therefore we did not try to incorporate it into our theoretical calculations (using a pentalayer FSFSF structure). Still, the non-monotonicity of our data around $P[$Ni$]/(1-P[$Ni$]) = 0.2$ might be another indication for the role played by spin-triplet superconductivity in our samples. While the claim for non-monotonicity is not conclusive, we were unable to find a single set of parameters in the bilayer model that simultaneously fits our data in the low relative volume concentrations regime of Ni/Pb and in the high relative volume concentration regime. 

\section{CONCLUSIONS}
In conclusion, in this work we examined the superconducting transition in SF Pb-Ni granular films with varying composition. We have fitted them to approximate SF layered models, and found that with reasonable values of the parameters, an agreement could only be obtained with a trilayer model, but not with a bilayer one. This hints at the importance of triplet Cooper-pairing in our samples, which results from the variation of the magnetization direction between different Ni grains.

\section{ACKNOWLEDGEMENTS}
The theoretical research of this work was supported by the Israel Science Foundation (ISF; Grant 227/15), the German-Israeli Foundation (GIF; Grant I-1259-303.10), the US-Israel Binational Science Foundation (BSF; Grant 2014262), and the Israel Ministry of Science and Technology (MOST; Contract 3-12419).

\appendix*
\section{SOLUTION OF THE USADEL EQUATIONS FOR FSF TRILAYERS}
In this Appendix we describe the analytical and numerical scheme devised by L\"ofwander \textit{et al.}~\cite{lofwander} for the solution of the Usadel equations~(\ref{eq:UsadelEquations})--(\ref{eq:boundaryconditions}) of a FSF trilayer, which we have employed in our calculations.
In the superconducting region, the two Usadel equations [Eq.~(\ref{eq:UsadelEquations})] are decoupled, and the
triplet component can be solved for analytically:
\begin{equation}
\vec{f}_{t}(x) =  \vec{c} \cosh(k_{s}x)+\vec{d} \sinh(k_{s}x),
\label{eq:tripletpart}
\end{equation}
where $k_{s} = \sqrt{2|\epsilon_{n}|/D_{s}}$.
The presence of the ferromagnetic regions is reduced to an effective
boundary condition for the calculation of the singlet component in
the superconducting region, 
$0<x<d_{s}$:
\begin{equation}
\left(\begin{array}{c}
\partial_x f_{s}(0)\\
\partial_x f_{s}(d_{s})
\end{array}\right) = k_{s}
\left(
\begin{matrix}
W_{11} & W_{12} \\
W_{21} & W_{22}
\end{matrix}
\right)
\left(\begin{array}{c}
f_{s}(0)\\
f_{s}(d_{s})
\end{array}\right),
\end{equation}
where the coefficients $W_{ij}$ are determined by plugging the
solutions for $f_{s}$ into the original boundary conditions, Eq.~(\ref{eq:boundaryconditions}):
\begin{eqnarray}
W_{11} & = & \frac{C_{+}\left(B_{+}I_{-,-}+B_{-}I_{+,-}\right)+C_{-}\left(B_{+}I_{-,+}+B_{-}I_{+,+}\right)}{J}, \nonumber \\
W_{22} & = & -\frac{D_{+}\left(A_{+}I_{-,-}+A_{-}I_{-,+}\right)+D_{-}\left(A_{+}I_{+,-}+A_{-}I_{+,+}\right)}{J}, \nonumber \\
W_{12} & = & \frac{2K_{0}^{2}\cos^{2}\theta\left(B_{-}D_{+}-B_{+}D_{-}\right)\left(A_{-}C_{+}-A_{+}C_{-}\right)}{J}, \nonumber \\
W_{21} & = & -W_{12},
\end{eqnarray}
with:
\begin{eqnarray}
A_{j} & = & \cosh\left(k_{j1}d_{f1}\right)+\gamma_{b1}k_{j1}\xi_{f1}\sinh\left(k_{j1}d_{f1}\right),\nonumber \\
B_{j} & = & \cosh\left(k_{j2}d_{f2}\right)+\gamma_{b2}k_{j2}\xi_{f2}\sinh\left(k_{j2}d_{f2}\right),\nonumber \\
C_{j} & = & \gamma_{1}k_{j1}\xi_{f1}\sinh\left(k_{j1}d_{f1}\right)/k_{s}\xi_{s},\\
D_{j} & = & \gamma_{2}k_{j2}\xi_{f2}\sinh\left(k_{j2}d_{f2}\right)/k_{s}\xi_{s},\nonumber \\
K_{j} & = & \left(B_{j}C_{0}+D_{j}A_{0}\right)\cosh\left(k_{s}d_{s}\right)+\left(B_{j}A_{0}+D_{j}C_{0}\right)\sinh\left(k_{s}d_{s}\right), \nonumber
\end{eqnarray}
for $j=0,\pm$, with
$k_{\pm i} = \sqrt{\left(2\epsilon_{n}\pm2iJ_{i}\right)/D_{fi}}$,
$k_{0i} = \sqrt{2\epsilon_{n}/D_{fi}}$, and $\theta$ being the angle between the magnetization vectors of the two ferromagnetic layers.
In addition:
\begin{widetext}
\begin{equation}
J = A_{+}B_{+}I_{-,-}+A_{-}B_{-}I_{+,+}+A_{+}B_{-}I_{+,-}+A_{-}B_{+}I_{-,+},
\end{equation}
where:
\begin{eqnarray}
I_{\epsilon\epsilon'} & = & -2K_{0}\cos\theta[\left(K_{\epsilon}B_{0}\sin^{2}\theta+K_{0}B_{\epsilon}\cos^{2}\theta\right)\left(A_{\epsilon'}\sinh\left(k_{s}d_{s}\right)+C_{\epsilon'}\cosh\left(k_{s}d_{s}\right)\right) \nonumber\\
 &  & +\left(K_{\epsilon}D_{0}\sin^{2}\theta+K_{0}D_{\epsilon}\cos^{2}\theta\right)\left(C_{\epsilon'}\sinh\left(k_{s}d_{s}\right)+A_{\epsilon'}\cosh\left(k_{s}d_{s}\right)\right)].
\end{eqnarray}
\end{widetext}

In order to calculate $T_{c}$, we can consider the first Usadel equation in
the superconducting region, where $\vec{J}=0$, and obtain:
\begin{equation}
f_{s}(\epsilon_{n},x) = \pi\intop_{0}^{d_{s}}G(\epsilon_{n},x,y)\Delta(y)dy,
\end{equation}
in terms of the appropriate Green function $G(\epsilon_{n},x,y)$. The gap equation then assumes the form:
\begin{equation}
\frac{2\pi T\sum_{\epsilon_{n}>0}\int_{0}^{d_{s}}G(\epsilon_{n},x,y)\Delta(y)dy}{\ln\frac{T}{T_{cs}}+2\pi T\sum_{\epsilon_{n}>0}(\epsilon_{n})^{-1}} = \Delta(x).
\end{equation}
Lofwander \textit{et al.}\ proposed to solve this equation using a spatial Fourier series expansion of $\Delta(x)$
\begin{equation}
\Delta(x) = \sum_{p=0}^{\infty}\Delta_{p}\cos\left(\frac{p\pi x}{d_{s}}\right),
\end{equation}
where the coefficients $\Delta_{p}$ are:
\begin{equation}
\Delta_{p} =  \frac{2-\delta_{p0}}{d_{s}}\int_{0}^{d_{s}}\Delta(x)\cos\left(\frac{p\pi x}{d_{s}}\right)dx.
\end{equation}
Consequently, in the Fourier coefficient space, the gap equation is
written as:
\begin{eqnarray}
\sum_{p=0}m_{lp}\Delta_{p} & = & 0,
\label{eq:trilayerequation}
\end{eqnarray}
for integer $l\geq0$, where 
\begin{eqnarray}
m_{lp} & = & 4\pi T\sum_{\epsilon_{n}>0}\frac{1}{\epsilon_{n}}b_{lp}\beta_{l}\beta_{p},\label{eq:D3}\\
m_{ll} & = & (1+\delta_{l0})\ln\frac{T}{T_{cs}}+4\pi T\sum_{\epsilon_{n}>0}\frac{1}{\epsilon_{n}}\left[b_{ll}\beta_{l}^{2}+\frac{1}{2}(1-\beta_{l})\right],\nonumber 
\end{eqnarray}
with $b_{lp}$ given by: 
\begin{widetext}
\begin{equation}
b_{lp} = 
\frac{\left[W_{11}-\left(-1\right)^{l+p}W_{22}+\left(-1\right)^{p}W_{12}-\left(-1\right)^{l}W_{21}\right]\sinh\left(k_{s}d_{s}\right)+\det\left(W\right)\left\{ \left(-1\right)^{p}+\left(-1\right)^{l}-\left[1+\left(-1\right)^{l+p}\right]\cosh\left(k_{s}d_{s}\right)\right\} }{k_{s}d_{s}L},
\end{equation}
in which
\begin{equation}
L = W_{12}-W_{21}+\left(W_{11}-W_{22}\right)\cosh\left(k_{s}d_{s}\right)+\left[1-\det\left(W\right)\right]\sinh\left(k_{s}d_{s}\right).
\end{equation} 
\end{widetext}

Eq.~(\ref{eq:trilayerequation}) 
can be solved numerically by introducing a cutoff $p_{c}$
for the number of harmonics, and finding the critical temperature as
the highest temperature for which the eigenvalue of the $p_{c} \times p_{c}$
matrix $m_{lp}$ equals 0.  In order to speed up the numerical calculations
one may separate the matrix $m_{lp}$ into two terms:
\begin{eqnarray}
m_{lp} & = & \bar{m}_{lp}+R_{lp},
\end{eqnarray}
where $\bar{m}_{lp}$ includes the sum in Eq.~(\ref{eq:D3}) up to a cutoff $\epsilon_{c}$, while the term $R_{lp}$ is the sum from $\epsilon_{c}$
to infinity, which is approximated by an integral:
\begin{eqnarray}
R_{lp} & = & \delta_{lp}\frac{1}{\pi}\ln\left(1+\frac{p^{2}}{\widetilde{d_{s}}^{2}}\frac{T_{c0}}{\epsilon_{c}}\right)+\frac{2}{\pi^{2}}\frac{1}{\widetilde{d_{s}}}I_{lp}\label{eq:F10}\\
I_{lp} & = & \int_{\epsilon_{c}/T_{c0}}^{\infty}\frac{c_{1}\sqrt{x}+c_{2}x}{\left(x+\frac{l^{2}}{\widetilde{d_{s}}^{2}}\right)\left(x+\frac{p^{2}}{\widetilde{d_{s}}^{2}}\right)\left(c_{3}+c_{4}\sqrt{x}+c_{5}x\right)}dx.
\nonumber
\end{eqnarray}

\bibliographystyle{apsrev4-1}
\bibliography{articlebib}

\end{document}